\documentclass[twocolumn,prb,showpacs,superscriptaddress,groupedaddress]{revtex4-1}
\usepackage{color}
\usepackage{amssymb}
\usepackage{dcolumn}
\usepackage{amstext}
\usepackage{graphicx}
\usepackage{amsmath}
\usepackage[normalem]{ulem}
\usepackage{dsfont}

\newcommand{\be}{\begin{equation}}
\newcommand{\ee}{\end{equation}}
\newcommand{\ba}{\begin{eqnarray}}
\newcommand{\ea}{\end{eqnarray}}

\newcommand{\vk}{{\mathbf{k}}}

\newcommand{\ut}[1]{\mathrm{\; #1}}

\graphicspath{{./figures/}}

\begin{document}

\title{Using controlled disorder to distinguish $s_\pm$ and $s_{++}$ gap structure
 in Fe-based superconductors}

\author{Y.~Wang}
\author{A.~Kreisel}
\author{P.J.~Hirschfeld}
\affiliation{Department of Physics, University of Florida, Gainesville, Florida 32611, USA}

\author{V.~Mishra}
\affiliation{Materials Science Division, Argonne National Laboratory, Lemont, Illinois 60439, USA}

\date{\today}

\begin{abstract}
We reconsider the effect of disorder on the properties of a superconductor characterized by a sign-changing
order parameter appropriate for Fe-based materials.  Within a simple two band model, we calculate
simultaneously $T_c$, the change in residual resistivity $\Delta \rho_0$, and the zero-energy density of
states, and show how these results change for various types of gap structure and assumptions regarding the
impurity scattering. The rate of $T_c$ suppression is shown to vary dramatically according to details of the
impurity model considered. We search therefore for a practical, experimentally oriented signature of a gap of
the $s_\pm$ type, and propose that observation of a particular evolution of the penetration depth, nuclear
magnetic resonance relaxation rate, or thermal conductivity temperature dependence with disorder would
suffice.
\end{abstract}


 \maketitle

\section{Introduction}
Determining the symmetry and structure of the superconducting order parameter in iron-based superconductors
(FeSCs) is one of the main challenges in this new field.~\cite{Hirschfeld_rev11,*Chubukov12} The Fermi
surface is usually given by two or three $[\Gamma\!=\!(0,0)]$-centered hole pockets and two
$[M\!=\!(\pi,\pi)]$-centered electron pockets  in the two-Fe zone composed primarily of Fe $3d$ states.
Repulsive interband interactions between hole and electron pockets leading to spin fluctuations are often
assumed to lead to a superconducting order parameter which changes sign over the Fermi surface (FS) to lower
the overall Coulomb energy. The simplest version of this state, called the $s_{\pm}$~state, is described by
an isotropic order parameter on each FS with opposite signs for electronlike and holelike
pockets.~\cite{Mazin08} The state may be highly anisotropic and even exhibit gap nodes, but still be
considered $s_\pm$ provided the average sign on hole pockets is opposite that on electron pockets. On the
other hand, other theories suggest that orbital fluctuations may dominate the pairing interactions in systems
of this type, favoring a gap with equal sign on all pockets, denoted $s_{++}$.~\cite{Kontani10}

Surprisingly, it has proven rather difficult to definitively distinguish these types of gap structures
experimentally, in part because phase-sensitive experiments are challenging due to surface properties;
because of the multiband nature of the electronic structure; and because the $s_{\pm}$ and $s_{++}$ ``states"
are symmetry equivalent, transforming both according to the $A_{1g}$ representation of the crystal point
group. At this writing, three experiments offer indirect evidence in favor of the $s_\pm$~state: the nearly
ubiquitous observation of neutron spin resonance features in inelastic neutron spectroscopy
(INS),~\cite{Lumsden09,Christianson09,Inosov10,Park10,Argyriou10,Castellan11} a quasiparticle interference
scanning tunneling spectroscopy (STS) experiment in a magnetic field,~\cite{Hanaguri10} and a phase-sensitive
experiment on a polycrystalline sample which relies on significant statistical analysis.~\cite{Chen10}

On the other hand, alternative explanations have been offered for all these measurements; in particular,
Kontani and Onari have provided an alternate explanation~\cite{Kontani10} for the neutron resonance features
within an $s_{++}$ scenario via a postulated energy dependence of the quasiparticle relaxation time. In
addition, several references~\cite{Senga08,Senga09,Onari09,Lietal12,Kirshenbaum12} have called attention to a
``slow" decrease of $T_c$ in chemical substitution
experiments,~\cite{Li10,Nakajima10,Tropeano10,Lietal12,Kirshenbaum12} which is then ascribed to the natural
robustness  against nonmagnetic disorder of an $s_{++}$ superconductor. It is this issue which we study here.

It is important to understand what is meant by ``slow'' and ``fast'' $T_c$ suppression in this context.  At
one extreme we have situations in which $T_c$ is not suppressed by nonmagnetic disorder at all.  According to
Anderson's theorem, the critical temperature of an isotropic conventional $s$-wave superconductor with a
single band of electrons is unaffected by nonmagnetic scatterers.  From this statement it follows immediately
that   the same occurs for two bands in an isotropic $s_{++}$ state (with equal gaps), but also  in an
$s_\pm$~state with no interband scattering.  At the other extreme, we know that magnetic scatterers in a
conventional isotropic superconductor suppress $T_c$ according to the Abrikosov-Gor'kov (AG)
law;~\cite{Abrikosov61} it is well known that \textit{nonmagnetic} scatterers suppress $T_c$ at the same fast
AG rate in a two-band $s_\pm$~state, \textit{provided}  the two densities of states $N_a=N_b$ and two gaps
$\Delta_a=-\Delta_b$ are equal in magnitude, and the scattering is purely \textit{interband} in nature. Any
deviation from these assumptions will \textit{slow} the $T_c$ suppression rate relative to the AG rate.
Therefore between these two extremes lie many possibilities for $T_c$ suppression behavior which depend on
details of the electronic structure and the relative amplitudes of inter- and intraband scattering.

Several theoretical calculations of $T_c$ suppression  have discussed the pairbreaking effects of nonmagnetic
scatterers on model multiband superconductors with generalized $s$-wave
order.~\cite{Preosti96,Golubov97,*Golubov95,Parker08,Chubukov08,Senga08,*Senga09,Bang09,Mishra09,Golubov02,Kogan09,Kulic99,*Ohashi04,Efremov11}
In fact the situation is generally even more complicated than discussed above or in these works, since
chemical impurities may do more than simply provide a scattering potential: they may dope the system, or
alter the pairing interaction itself locally.  We therefore believe (see also
Ref.~\onlinecite{Hirschfeld_rev11,*Chubukov12}) that measurements of $T_c$ suppression relative to the amount
of chemical disorder are not particularly useful to determine the gap structure in multiband systems. To
improve the situation, one first needs to find a way to create pointlike potential scattering centers, so as
to create  disordered systems to which the above theoretical works apply. The closest approach to this ideal
is achieved with low-energy electron irradiation, which is thought to create interstitial-vacancy pairs.
Experiments of this type are being performed currently, and it is one of the goals of this work to make
predictions to guide the analysis of such data.

The other needed improvements are theoretical: first, the pairbreaking theory must be extended to relate
$T_c$ only to directly measurable quantities, like the change in residual ($T\rightarrow 0$)  resistivity
caused by the disorder, rather than to any theoretically meaningful but empirically inaccessible scattering
rate parameter.  Second, since the theory involves many parameters, the robustness of any claimed fit must be
tested by the simultaneous prediction of other quantities which depend on disorder, such as the
low-temperature penetration depth, nuclear magnetic resonance (NMR) relaxation rate, or thermal conductivity.
Finally, it would be useful to have \textit{ab initio} calculations of vacancy and interstitial potentials to
constrain the impurity parameters used.  This has been attempted for chemical
substituents~\cite{Kemper09,*Kemper10_erratum, Nakamura11} recently.

\section{Model}
\begin{figure}[!b]
  \includegraphics[width=1.0\linewidth]{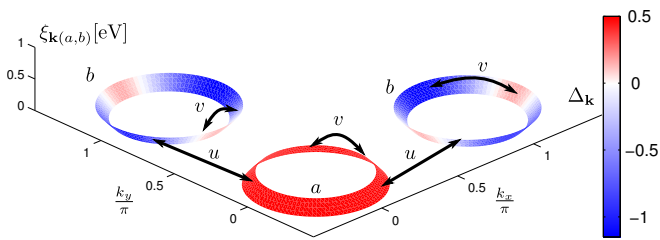}
\caption{(Color online) Sketch of the two-band model with linearized band dispersions on the Fermi sheets $a$
and $b$ and constant impurity scattering $v$ (intraband) and $u$ (interband), together with a possible nodal
$s$-wave gap on the bands in the superconducting state.}
  \label{fig0}
\end{figure}

We consider a system with two bands $a$ and $b$ with linearized dispersion close to the Fermi level that lead
to densities of state $N_a$ and $N_b$ in the normal state; see Fig.~\ref{fig0}.

The $t$-matrix equation in the two-band model has the form
\begin{subequations}
 \begin{eqnarray}
   \mathbf{\hat{\Sigma}}  &=& {n}_\text{imp}\mathbf{\hat{t}},  \label{t2b} \\
   \mathbf{\hat{t}}       &=& \mathbf{\hat{u}}+\mathbf{\hat{u}\hat{g}} \mathbf{\hat{t}},  \label{tmat}
 \end{eqnarray}
\end{subequations}
where $n_{\mathrm{imp}}$ is the concentration of impurities,
$\mathbf{\hat{t}}(n_{\mathrm{imp}})=\sum_{i=0}^{3}\mathbf{t}^{(i)}\mathbf{\otimes}\hat{\tau}_{i}$,
$\mathbf{\hat{g}}(n_{\mathrm{imp}})=\mathbf{g}_{0}\mathbf{\otimes}\hat{\tau}_{0}+\mathbf{g}_{1}\mathbf{\otimes}\hat{\tau}_{1}$
and $\mathbf{\otimes}$ represents a product of band (bold) and Nambu (caret) matrices.
$\mathbf{g}_{0}=\mathop\mathrm{diag}(g_{0a},g_{0b})$ and $\mathbf{g}_{1}=\mathop\mathrm{diag}(g_{1a},g_{1b})$
are local Green's functions in the $\tau_0$ and $\tau_1$ channels (we have assumed particle-hole symmetry in
order to neglect $\mathbf{g}_{3}$), where $\hat\tau_i$ denote Pauli matrices in Nambu space. Due to the
translational invariance of the disorder-averaged system, $\mathbf{\hat{g}}$ is diagonal in band space.  We
now assume a simple model for impurity scattering whereby electrons scatter within each band with amplitude
$v$ and between bands with amplitude $u$,
\begin{equation}
  \hat{{\bf u}}=\left( \begin{array}{cc} v & u \\ u & v \label{eq:uN} \end{array} \right)
  \mathbf{\otimes}\hat{\tau}_{3}.
\end{equation}
The $t$-matrix components are found from Eq.~(\ref{tmat}) to be
\begin{eqnarray}
 t_{aa}^{(0)} &=&  \frac{\left[ g_{0b}u^{2}+g_{0a}v^{2}-g_{0a}
    \left( u^{2}-v^{2}\right)^{2}\delta g_b^2\right]}{\mathcal D}, \nonumber \\
 t_{aa}^{(1)} &=& -\frac{\left[ g_{1b}u^{2}+g_{1a}v^{2}-g_{1a}
    \left( u^{2}-v^{2}\right)^{2}\delta g_b^2\right]}{\mathcal D},
 \label{Sgen}
\end{eqnarray}
and $t_{bb}^{(i)}=t_{aa}^{(i)}(a\leftrightarrow b)$, where
\begin{eqnarray}
 {\mathcal D} &=& 1- \left(\delta g_a^2+\delta g_b^2\right) v^{2}+ \delta g_a^2\delta g_b^2
    \left(u^{2}-v^{2}\right) ^{2}-  \nonumber \\
    && -2u^{2}\left(g_{0a}g_{0b}-g_{1a}g_{1b}\right)
    \label{DD}
\end{eqnarray}
with the abbreviation $\delta g_\alpha^2=g_{0\alpha}^{2}-g_{1\alpha}^{2}$.

\section{$T_c$ suppression}
The linearized multiband gap equation near $T_{c}$ is (see, e.g. Ref.~\onlinecite{Mishra09})
\begin{equation}
\Delta_{\alpha}{(\mathbf{k})} = 2 T \!\sum_{{\bf k}^{\prime},\beta,\omega_{n} > 0}^{\omega_{n} = \omega_{c}}\!
V^{\alpha \beta}_{{\bf k} {\bf k}^{\prime}} \frac{\tilde{\Delta}_{\beta}({\bf
k}^{\prime})}{\tilde{\omega}^{2}_{\beta}+\xi^{2}_{\beta}}, \label{eq:gpetc}
\end{equation}
where $\xi_{\beta}$ is the linearized dispersion of band $\beta$, and we introduced the shifted gaps and
frequencies, $\tilde{\Delta}_{\beta}({\bf k}^{\prime})={\Delta}_{\beta}({\bf k}^{\prime})+\Sigma_\beta^{(1)}$
and ${\tilde{\omega}_{\beta}}=\omega_n+{i\Sigma_{\beta}^{(0)}}$. We will simplify the model above further in
that we adopt a gap structure similar to that obtained from spin fluctuation theories: The gap on the (hole)
pocket $a$ is isotropic, $\Delta_a$, and the gap on the (electron) pocket $b$ may be anisotropic,
$\Delta_b=\Delta_b^0 +\Delta_b^1(\theta)$, where $\theta$ is the momentum angle around the $b$ pocket and
$\int d\theta \Delta_b^1(\theta)=0$. The pairing potential is then taken as
$V^{\alpha\beta}_{\vk\vk'}=V_{\alpha\beta}\phi_\alpha(\vk)\phi_\beta(\vk')$, with $\phi_\alpha=1
+r\delta_{\alpha,b}\cos 2\phi$, and $\phi$ is the angle around the electron pocket. The parameter $r$
controls the degree of anisotropy, and creates nodes if $r>1$.

This ansatz then gives three coupled gap equations for {$(\Delta_a,\Delta_b^0,\Delta_b^1)^\text{T}\equiv
\underline{\Delta}$}. In the {$\underline{\Delta}$} basis  we can write the gap equations in the compact form
\begin{equation}
 \underline{\Delta}={\ln}\Bigl(1.13\frac{\omega_{c}}{T_{c}}\Bigr) \underline{\mathcal{M}}\,\underline{\Delta}
  {\equiv \mathcal{L}_0\underline{\mathcal{M}}\,\underline{\Delta}},
\label{eq:tcc1}
\end{equation}
the matrix
$\underline{\mathcal{M}}=({1}+\underline{V}\,\underline{\mathcal{R}}^{-1}\underline{\mathcal{X}}\,\underline{\mathcal{R}})^{-1}\,\underline{V}$
and the constant $\mathcal{L}_0=\ln\bigl(1.13\frac{\omega_{c}}{T_{c}}\bigr)$ were introduced. Here
$\underline{V}$ is the interaction matrix in the above basis. $\underline{\mathcal{R}}$ is the orthogonal
matrix which diagonalizes the matrix $\underline{\Lambda}$,
\begin{equation}
 \underline{\Lambda}=\frac{\pi n_\text{imp}}{\mathcal D_N} \left[\begin{array}{ccc}
                      N_{b}u^{2} & {-}N_{b}u^{2} & 0 \\
                      {-}N_{a}u^{2} & N_{a}u^{2} & 0 \\
                      0 & 0 & {N_{b}v^{2}+N_au^2} \\
                     \end{array}\right],
\label{eq:Lambda}
\end{equation}
where
\begin{equation}
\mathcal D_N=1+2u^2\pi^2N_aN_b +(u^2-v^2)^2\pi^4N_a^2 N_b^2 +v^2\pi^2(N_a^2+N_b^2)
\end{equation}
is Eq.~(\ref{DD}) evaluated in the normal state where the limit $\underline \Delta \rightarrow 0$ has been
taken in the local Greens functions. $\underline{\mathcal{X}}$ is a matrix with only diagonal elements,
\begin{equation}
{\mathcal{X}}_{ii}=
{ \mathcal{L}_0-\left[\Psi\Bigl(\frac{1}{2}+\frac{\omega_c}{2\pi T_c}+\frac{\lambda_{i}}{2\pi
T_c}\Bigr)-\Psi\Bigl(\frac{1}{2}+\frac{\lambda_{i}}{2\pi T_c}\Bigr) \right]} \label{eq:Xd},
\end{equation}
where $\Psi$ is the digamma function and  $\lambda_{i}$ are the eigenvalues of the matrix
$\underline{\Lambda}$. The maximum eigenvalue $[\lambda_\text{max}(T_c)]$ of the matrix
$\underline{\mathcal{M}}$ determines $T_c$ via $
T_{c}=1.13\;\omega_{c}\exp\left[-1/\lambda_\text{max}(T_c)\right]$.

\section{Residual resistivity}
The most direct  observable measure of scattering in $T_c$ suppression experiments is the residual
resistivity change $\Delta \rho_0$,  i.e., the change in the extrapolated $T\rightarrow 0$ value of the
resistivity with disorder.  We will assume that interference effects between elastic and inelastic processes
are negligible, i.e., that the effect on the $\rho(T)$ curve when the system is disordered is essentially a
$T$-independent shift upward.  We therefore  calculate $\Delta \rho_0$  within the same framework as above,
assuming that all defects are pointlike. In the zero frequency limit, there are no interband transitions, and
the total conductivity in the $x$ direction is the sum of the Drude  conductivities of the two bands,
$\sigma=\sigma_a+\sigma_b,$ with $\sigma_\alpha=2e^2N_{\alpha} \langle v_{\alpha,x}^2\rangle \tau_\alpha$,
where $v_{\alpha,x}$ is the component of the Fermi velocity in the $x$ direction and $\tau_\alpha$ the
corresponding single particle relaxation time obtained from the self-energy in the $t$-matrix approximation,
$\tau_\alpha^{-1}=-2\mathop{\mathrm {Im}}\Sigma_{\alpha}^{(0)}$. Note that $\tau_\alpha^{-1}$ contains
contributions from both the intraband and interband impurity scattering processes.  The transport time and
single-particle lifetime are identical within this model because of our assumption of pointlike $s$-wave
scatterers, which implies that corrections to the current vertex vanish.  A finite spatial range of the
scattering potential will tend to steepen the $T_c$ vs $\Delta \rho_0$ curve.\cite{Graser_forward,Zhu04}
  \vskip .2cm

\section{Results} \label{sec_results}
\subsection{$T_c$ suppression vs resistivity}
We now solve Eqs. (\ref{eq:tcc1}) for $T_c$ and calculate simultaneously the change in resistivity $\Delta
\rho_0$ at $T\rightarrow 0$.  Unlike $T_c$ vs $n_\mathrm{imp}$ or various scattering rates, $T_c$ vs
$\Delta\rho_0$ can then be compared directly to experiment. Clearly, the results will be parameter dependent,
however, so we here specify our precise assumptions regarding the electronic structure.  For concreteness, we
focus on the BaFe$_2$As$_2$ (Ba122) system on which the largest number of measurements have been reported,
and give parameters for this system and corresponding references in the Appendix.

Using these parameters, we obtain for the isotropic case ($r=0$)  the zero temperature gap values of
$\Delta_{a0}^0=-1.79 T_{c0}$ and $\Delta_{b0}^0=1.73 T_{c0}$, whereas for the nodal case ($r=1.3$) these are
$\Delta_{a0}^0=-1.22 T_{c0}$ and $\Delta_{b0}^0=1.23 T_{c0}$ with the critical temperature chosen as
$T_{c0}=30$ K. We have fixed the intraband scattering potential at an intermediate strength value of
$v=0.25$, but show results for other values in the Appendix.  Potentials are given in eV and we set
$\hbar=k_\text{B}=1$.

\begin{figure}[!t]
  \includegraphics[width=1.0\linewidth]{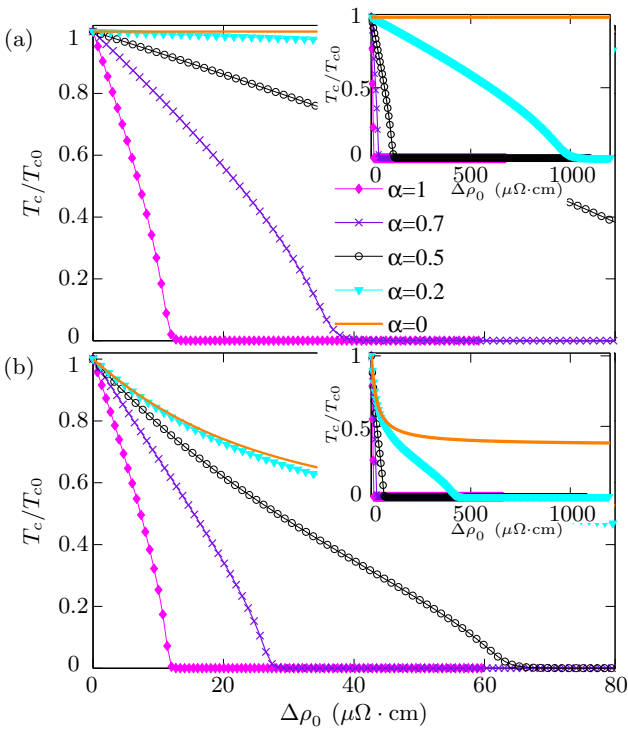}
\caption{(Color online) (a) Normalized critical temperature $T_c/T_{c0}$ vs disorder-induced resistivity
change $\Delta\rho_0$ for isotropic $s\pm$-wave pairing for various values of the inter- to intraband
scattering ratio $\alpha\equiv u/v$. Inset: Same quantity plotted over a larger $\Delta\rho_0$ scale.  (b) As
(a) but for an anisotropic (nodal) gap with anisotropy parameter $r=1.3$.}
  \label{fig1}
\end{figure}

In Fig.~\ref{fig1}, we now exhibit $T_c$ suppression vs the corresponding change in residual resistivity
$\Delta \rho_0$ as defined above, both for  a fully isotropic $s_\pm$~gap ($r=0$), and for a gap which has
nodes on the electron pockets ($r=1.3$), for a range of ratios $u/v$. It is clear that a wide variety of
initial slopes and critical resistivities $\Delta\rho_0^c$ for which $T_c\rightarrow 0$ is possible,
depending on the scattering character of the impurity. The variability of the suppression rate with the ratio
of inter- to intraband scattering has been noted by various authors\cite{Mishra09,Efremov11} before this. In
fact, Efremov \textit{et al.}\cite{Efremov11} have shown that the various $T_c$ suppression curves of the
isotropic $s_\pm$~gap fall onto universal curves depending on whether the average pair coupling constant
$\langle \lambda \rangle <,=,>0$ when plotted against the interband scattering rate (which is not directly
measurable, however). {Other works have made comparisons with the resistivity changes (for example Refs.
\onlinecite{Lietal12,Kirshenbaum12}), but have typically presented results for $s_{\pm}$ states only for a
single set of impurity parameters corresponding to the fastest rate of $T_c$ suppression.}  Such assumptions
lead always to critical $\Delta \rho_0$'s comparable to the smallest ones seen in Fig.~\ref{fig1}, of order
tens of $\mu\Omega$~cm.  Here we see that more general values of the parameters can easily lead to much
slower $T_c$ vs $\Delta \rho_0$ suppression rates by disorder, with critical disorder values of
$\Delta\rho_0^c$ of order m$\Omega$~cm.  As discussed by Li \textit{et al.},\cite{Lietal12} such values are
typical of chemical substitutions on various different lattice sites; here we see that such slow $T_c$
suppression does not rule out the $s_\pm$~state, even within the assumptions of our potential-scattering-only
model.

\subsection{Density of states}
A real understanding of the effects of disorder in a given situation will probably depend on correlating the
results of several experiments.  Other quantities which are quite sensitive to disorder are the temperature
dependence of the low-$T$ London penetration depth $\Delta \lambda(T)$ and the nuclear magnetic spin-lattice
relaxation time $T_1^{-1}$.  Within BCS theory, these quantities are controlled by the low-energy density of
states.  In the pure system, the nodal structure then determines the power law of temperature, and one
generically expects $\Delta\lambda(T)\sim T$ for gap line nodes except in very special
situations.\cite{Grossetal86}  In the presence of a small amount of nonmagnetic disorder, a finite density of
states is created \cite{GorkovKalugin,RiceUeda} which leads automatically to a $T^2$ term in the penetration
depth.\cite{Grossetal86,felds} If the state is of $s$ character, the gap nodes are not symmetry protected and
can be lifted by further addition of disorder.\cite{Borkowski,Mishra09}

\begin{figure}[!t]
\hspace{-1cm}
 \includegraphics[width=0.8\linewidth]{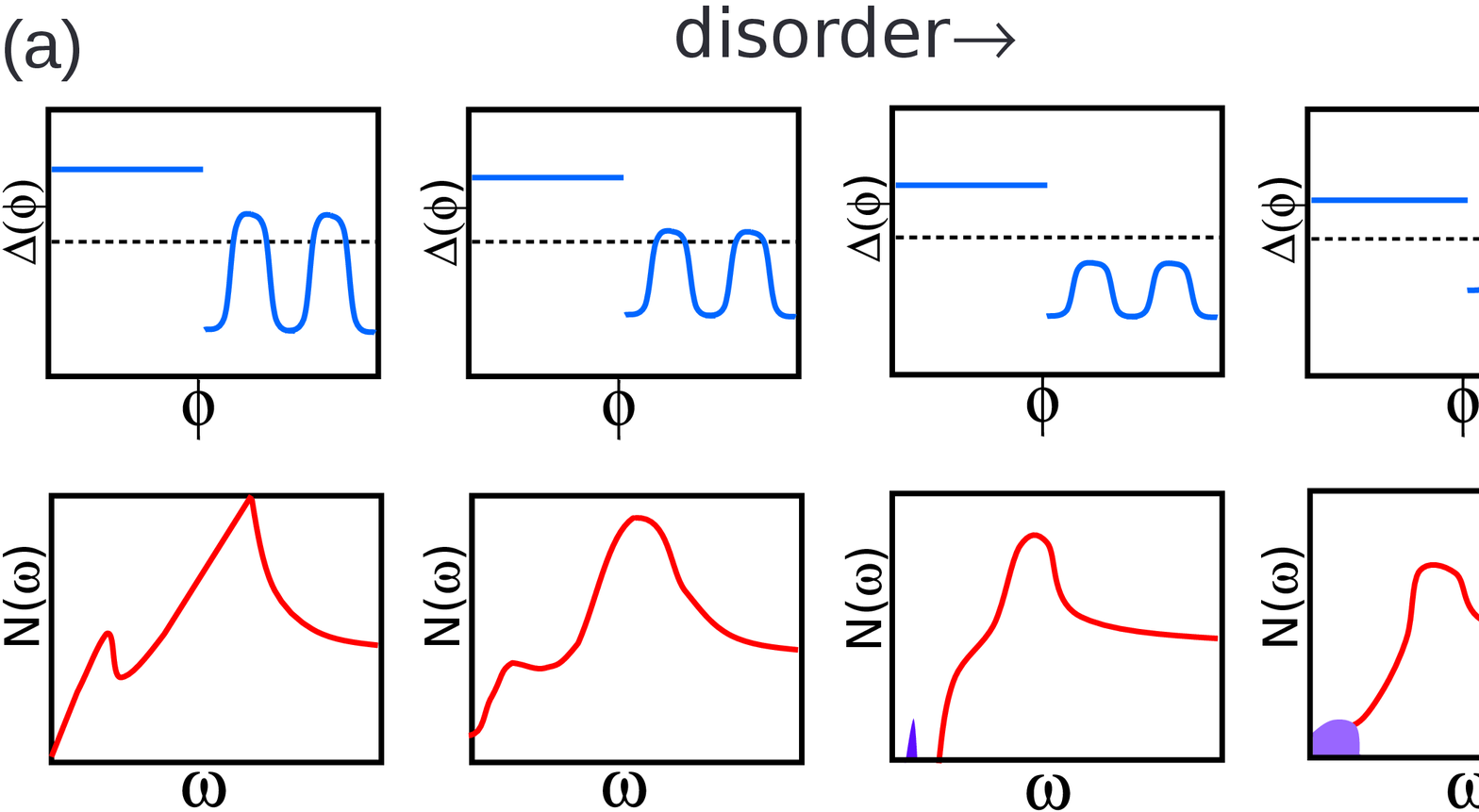}
 \includegraphics[width=0.9\linewidth]{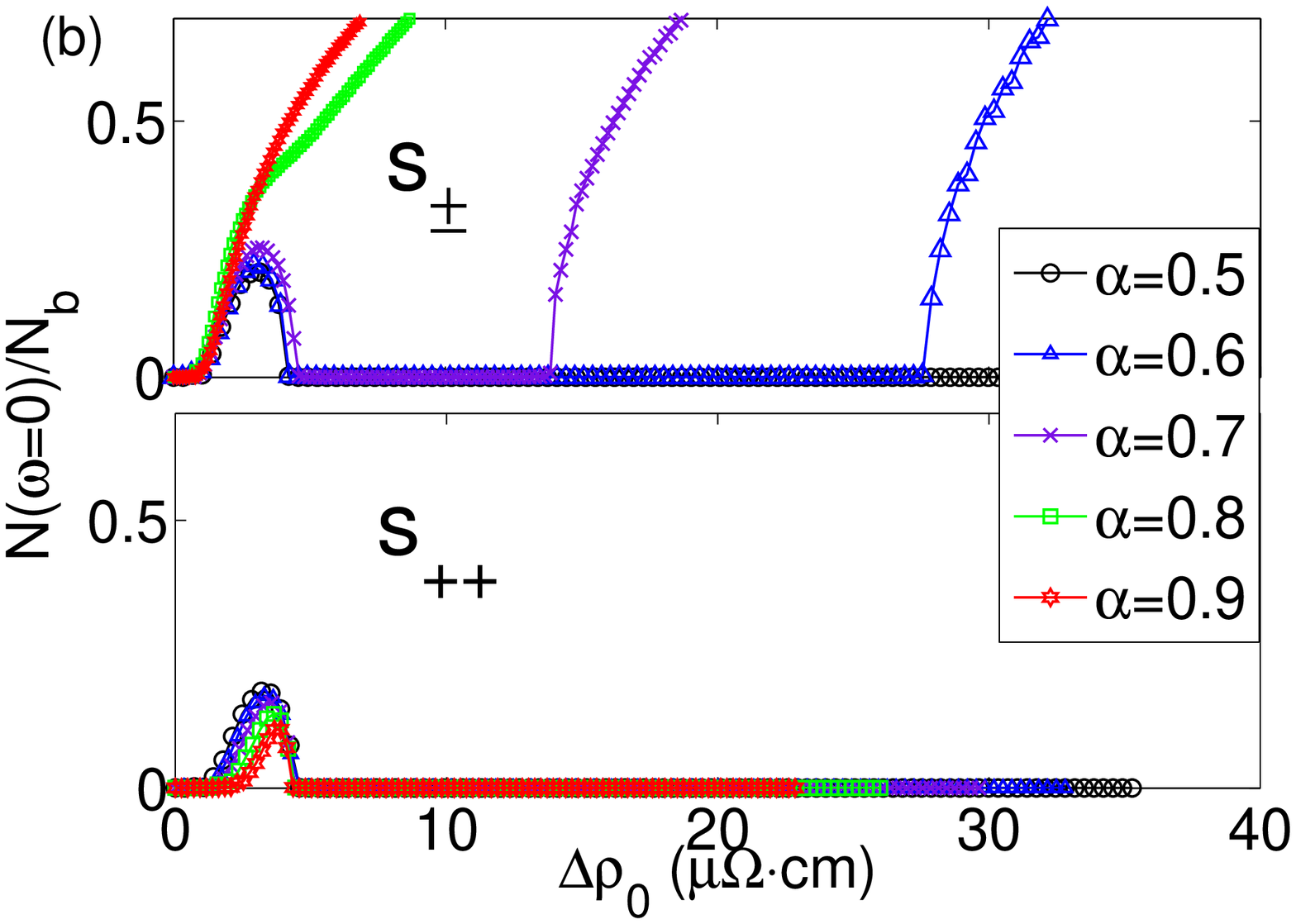}
\caption{(Color online) (a) Schematic evolution of the order parameter and density of states with increasing
disorder for a system with intra- and interband scattering.  (b) Top: Fermi level density of states $N_b(0)$
(nodal band) as shown in Fig.~\ref{fig1}(b) vs $\Delta \rho_0$ for various values of scattering ratio $u/v$
in an anisotropic $s_\pm$ state. Bottom: Fermi-level density of states for anisotropic $s_{++}$ state with
$V_{ab}$ identical in magnitude to the above panel, but positive. Anisotropy parameter $r=1.3$ in both
cases.} \label{fig2}
\end{figure}

In this work we note a further possibility in the disorder evolution of the low-energy density of states
(DOS) of a nodal multiband $s_\pm$-wave superconductor, namely, that a reentrant behavior of $N(0)$ can occur
after lifting of the nodes. The reason is that, in a situation dominated by intraband scattering but with
nonzero interband scattering, anisotropy of the gaps on each individual sheet will be averaged by intraband
disorder quickly.  If the state is $s_{\pm}$, a midgap impurity state can then be created by  interband
scattering, and grow until it overlaps the Fermi level, as shown schematically in Fig.~\ref{fig2} (a). Such
midgap states are the analogs of the Yu-Shiba bound states created by magnetic impurities in conventional
superconductors, and can appear for nonmagnetic impurities if the superconducting gap changes
sign.\cite{balatskyRMP} The residual density of states $N(0)=-\mathop{\mathrm{ Im}}\sum_{\bf
k}\mathop{\mathrm {Tr}} \hat G({\bf k},\omega=0)/(2\pi)$ ($\hat G$ is the Nambu Green's function) effectively
determines the low-energy thermodynamic behavior, so we have plotted it for the anisotropic band as a
function of increasing disorder in Fig.~\ref{fig2}, for both $s_\pm$ and $s_{++}$~states.  In the former case
the reentrant behavior is clearly seen.

The corresponding sequence in the $s_\pm$ penetration depth $\Delta \lambda(T)$ would be $T\rightarrow T^2
\rightarrow \exp(-\Omega_G/T) \rightarrow  T^2$, where $\Omega_G$ is the minimum gap in the system, while for
the NMR spin-lattice relaxation rate $T_1^{-1}$, the analogous evolution should be $T^3\rightarrow
T\rightarrow \exp(-\Omega_G/T) \rightarrow  T$. The residual linear $T$ term in the thermal conductivity,
$\kappa(T\rightarrow 0)/T$, should vanish and then reappear with increasing disorder.  In the $s_{++}$ case,
the last step in each sequence is entirely absent,  since interband scattering cannot give rise to low-energy
bound state formation.

\subsection{Realistic impurity potentials}
It is clear from the above analysis that we have established that there is a wide range of possibilities for
the behavior of $T_c$ in an $s_\pm$~superconductor, as well as for low-temperature properties like the
penetration depth, when disorder is systematically increased. To make more precise statements, one needs to
have some independent way to fix the scattering potential of a given impurity, and in particular the relative
proportion of inter- to intraband scattering. Kemper~\textit{et al.}~\cite{Kemper09} found the ratio between
inter- and intraband scattering to be of order $\alpha=0.3$ for Co in Ba122, which would lead according to
Fig.~\ref{fig1} to a critical resistivity strength of about {300 $\mu\Omega$~cm}, roughly in accord with
experiment.\cite{Lietal12,Kirshenbaum12}  Onari and Kontani\cite{Onari09} have made the important point that
the ``natural" formulation for a model impurity potential, i.e., diagonal in the basis of the five Fe $d$
orbitals, automatically leads to significant interband scattering if one transforms back to the band basis.
However, simple estimates show that depending on details $\alpha$ for on-site Fe substituents can vary
between 0.2 and 1, again leading as seen in Fig.~\ref{fig1} to a wide variety of possible $T_c$ suppression
scenarios.

\section{Conclusions}
We have argued that $s_\pm$~pairing cannot be ruled out simply because the $T_c$ suppression is slow
according to some arbitrary criterion.  The definitive experiments along these lines will most probably
involve electron irradiation, where one can be reasonably sure that the defects created act only as potential
scatterers.  In this case we find critical resistivities for the destruction of superconductivity which vary
over two orders of magnitude according to the ratio of interband to intraband scattering.   Results for the
$s_\pm$~state are then not inconsistent with experimental data, but proof of sign change of the order
parameter relies on knowledge of the impurity potential, which requires further \textit{ab initio}
calculations for each defect.  As an alternative approach, we have proposed that systematic variation of
disorder could give rise to a clear signature of $s_\pm$ pairing in the low-energy Fermi level DOS $N(0)$. In
an $s_\pm$ state, $N(0)$ could increase with disorder, vanish again due to node lifting, and increase again
afterward due to impurity bound state formation.  This ``reentrant" behavior of the DOS will be reflected in
the temperature dependence of low-temperature quasiparticle properties like the penetration depth, nuclear
spin relaxation time, or thermal conductivity. For some materials, this could be a ``smoking gun" experiment
for $s_\pm$~pairing.

\begin{acknowledgments}
The authors are grateful to C.J.~van der Beek, M.~Konczykowski, M.M.~Korshunov, R.~Prozorov,
F.~Rullier-Albenque, and T.~Shibauchi for discussions which stimulated the current project. P.J.H, Y.W., and
A.K. were supported by DOE Grant No. DE-FG02-05ER46236.  V.M.'s work  at Argonne National Laboratory, a U.S.
Department of Energy Office of Science Laboratory, was supported under Contract No. DE-AC02-06CH11357.
\end{acknowledgments}

\appendix*

\section{Model parameters}

\begin{figure}[!b]
 \includegraphics[width=1.0\linewidth]{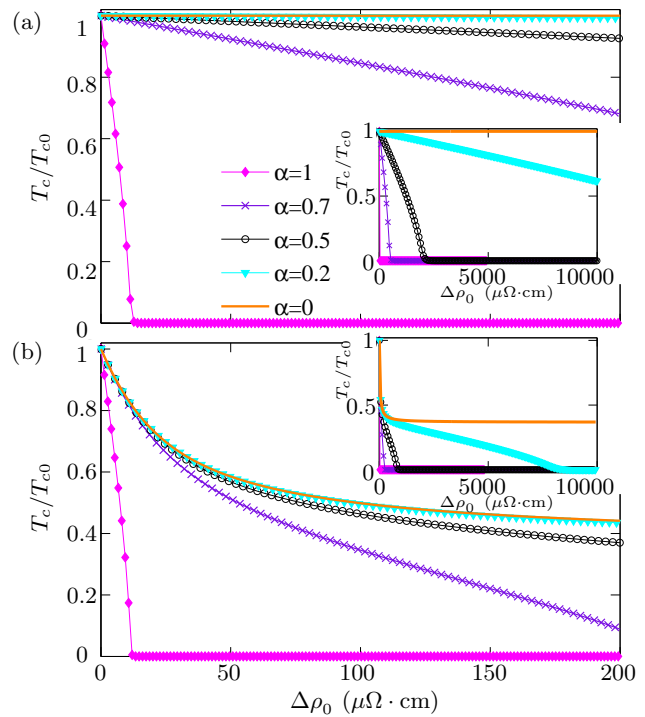}
\caption{(Color online) $T_c/T_{c0}$ vs. $\Delta\rho_0$  for various values of the inter- to intraband
scattering ratio $\alpha\equiv u/v$ with $v=1.25\ut{eV\,\mathcal{V}_c}$.  (a) for isotropic $s\pm$~wave
paring and (b) for an anisotropic (nodal) gap with anisotropy parameter $r=1.3$.}
 \label{fig:Tc-rho_v1p25}
\end{figure}

In this appendix we give some details of how our results change when taking values for the impurity
parameters and pair potential parameters different from those used in the main text, so that the reader may
judge how robust our conclusions are.

So far, we have focused on the parent compound BaFe$_2$As$_2$ and chosen values for the Fermi velocities and
densities of states at the Fermi level that are compatible with both density functional theory (DFT)
calculations \cite{Johnston10} and angle-resolved photoemission spectroscopy (ARPES)
measurements.\cite{Evtushinsky09,Ding11} {We assume a density of states on each Fermi surface sheet of
$N_a=3.6$ and $N_b=2.7/\mathcal{V}_c$/eV/spin ($\mathcal{V}_c$ is the unit cell volume), for the
``effective'' hole and electron pockets, respectively, that approximately describes the imbalance in the
densities of states that also has been seen with ARPES,\cite{Evtushinsky09,Ding11, Brouet09} and is
consistent with the density of states of Ba122 arising from Fe $d$-orbitals according to DFT
calculations\cite{Johnston10} with an effective-mass renormalization of $z=3$. We take the root-mean-square
Fermi velocities as $v_{\text{F},a}=2/3\times10^5$~m/s and $v_{\text{F},b}=10^5$~m/s from
Ref.~\onlinecite{Mishra11}, Table I, $v_{\text{F},\perp}$, and renormalize them by the same factor of $z=3$
to approximately match the velocities found in ARPES experiments.\cite{Evtushinsky09,Ding11, Brouet09}} In
the transport calculation, the component of the Fermi velocities in the direction of the current is taken to
be {$\langle v_{\text{F}\alpha,x}^2\rangle=1/2\, v_{\text{F}\alpha}^2$} due to the quasi-cylindrical Fermi
surface. The pairing potentials chosen for the main text are $V_{aa}=V_{bb}=0.05$ and $V_{ab}=V_{ba}=-0.04$.

\begin{figure}[tb]
 \includegraphics[width=1.0\linewidth]{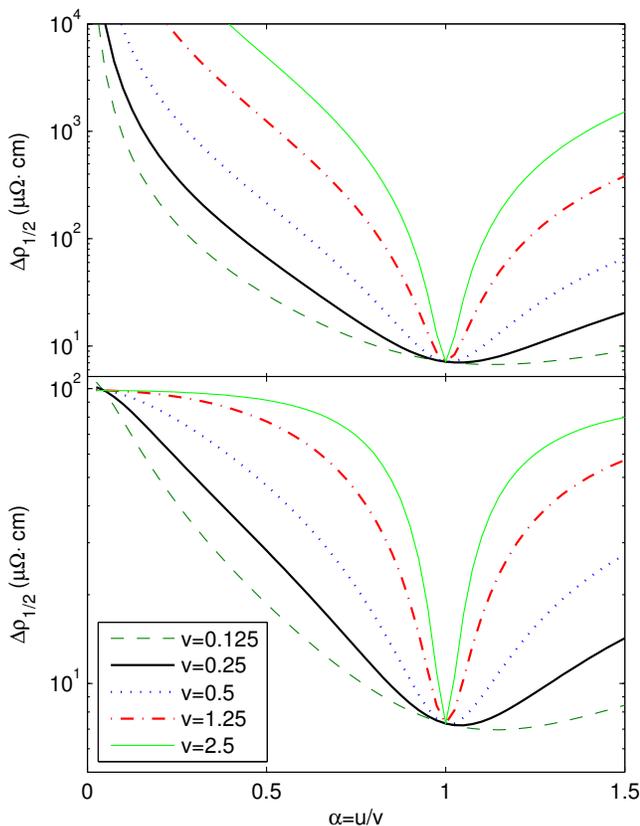}
\caption{(Color online) The resistivity at half suppression $\Delta\rho_{1/2}$ as a function of the ratio
$\alpha=u/v$ for various intraband impurity potentials $v$ (measured in $\text{eV}\,\mathcal{V}_c$); the
other parameters are taken as in the main article for the isotropic $s_{\pm}$~wave pairing (top) and for an
anisotropic (nodal) gap (bottom).}
 \label{fig_rho_12}
\end{figure}

However, there are still two parameters unfixed, namely, the pairing potential $V_{\alpha \beta}$ and the
impurity potential $v$ for scattering within bands (a full discussion of the variation of the inter- to
intraband potential ratio $\alpha=u/v$ is included in Sec.~\ref{sec_results}). Although the effective pairing
potential $V_{\alpha\beta}$ and average coupling constant $\langle \lambda
\rangle=\frac{1}{N_a+N_b}\sum_{\alpha,\beta \in \{a,b\}}N_\alpha V_{\alpha\beta}N_\beta$ as defined in
Ref.~\onlinecite{Efremov11} for our weak-coupling model, as well as the impurity scattering potentials $u$
and $v$,  are  not known in experiments, our conclusions are consistent with different parameters within a
reasonable range. If we increase $v$ to $v=1.25\ut{eV\,\mathcal{V}_c}$  keeping all other parameters
identical to those of Fig.~\ref{fig1} of the main text, the $T_c$ suppression significantly slows, as seen in
Fig.~\ref{fig:Tc-rho_v1p25}, with the exception of the  value $\alpha=1$, which plays a special role in the
theory of two-band $s_\pm$ superconductivity, as can be easily checked analytically. While in
Ref.~\onlinecite{Kontani10} it was argued that the interband scattering potential $u$ should be generically
large for any chemical substituent, there is no reason to expect $\alpha=1$ to hold exactly, and therefore we
see that large critical resisitvities $\Delta \rho_0^c$ are even more likely to be found for stronger
impurities (the unitarity limit $v\rightarrow \infty$ with fixed $\alpha$ is pathological in this
model\cite{Efremov11} and we have not considered it here). The special role of the value $\alpha=1$ can be
illustrated by plotting the resistivity $\Delta\rho_{1/2}$ at which the critical temperature is suppressed by
half, $T_c=0.5T_{c0}$, as shown in Fig.~\ref{fig_rho_12}, which may be compared with experiments. Note that
$\alpha\simeq 1$  yields the fastest $T_c$ suppression independent of the impurity potential in the physical
regime $v\gtrsim u$.

Finally, we  also mention the effect of choosing other pairing potentials $V_{\alpha\beta}$ that lead to
different values of $\langle \lambda\rangle$. As explained in Ref.~\onlinecite{Efremov11}, for isotropic
$s_\pm$ paring, when  $T_c$ is plotted vs. the effective interband  scattering rate, it follows three
different universal curves according to whether $\langle \lambda \rangle$ is greater than, equal to, or less
than 0. We have used a value $\langle \lambda \rangle =0.037\approx 0$ in our investigations. We have
examined other parameter sets with negative $\langle\lambda\rangle$, and found no essential difference in
$T_c$ {\it when plotted against the residual resistivity $\Delta\rho_0$,} which of course depends on both
intra- and interband scattering.



%

\end{document}